# LITTLE BLACK HOLES: DARK MATTER AND BALL LIGHTNING

MARIO RABINOWITZ
Armor Research, lrainbow@stanford.edu
Redwood City, CA 94062-3922, U.S.A.

**Abstract.** Small, quiescent black holes can be considered as candidates for the missing dark matter of the universe, and as the core energy source of ball lightning. By means of gravitational tunneling, directed radiation is emitted from black holes in a process much attenuated from that of Hawking radiation, $P_{SH}$, which has proven elusive to detect. Gravitational tunneling emission is similar to electric field emission of electrons from a metal in that a second body is involved which lowers the barrier and gives the barrier a finite rather than infinite width. Hawking deals with a single isolated black hole. The radiated power here is $P_R \propto e^{-2\Delta\gamma} P_{SH}$, where $e^{-2\Delta\gamma}$ is the transmission probability.

## 1. Introduction

Though black holes were long considered to be a fiction, their existence now seems to be firmly established. On an astronomical scale, black holes are the centers of attraction of galaxies. In our own galaxy and in the galaxy NGC 4258, the central dark mass is a black hole. In the case of our galaxy, recent measurements of the velocities of stars as close as 5 light days from the dynamical center imply a black hole of $2.6 \times 10^6$ solar masses (Genzel, 1998). Supermassive black holes of $10^6 - 10^{10}$ solar masses generate the vast power emitted by quasars, so that their luminosity far exceeds the luminosity of their entire galaxy (Davies, 1992). Trofimenko (1990) has discussed the possibility that little black holes are involved in a multitude of geophysical and astrophysical phenomena.

Zel'dovich's (1971) model of radiation from a rotating black hole is that "The rotating body produces spontaneous pair production [and] in the case when the body can absorb one of the particles, ... the other (anti)particle goes off to infinity and carries away energy and angular momentum." This is quite similar to the model later used by Hawking (1974, 1975) to propose radiation from non-rotating black holes. He also suggested that small black holes in stellar objects such as our sun might help to explain the solar neutrino problem (Hawking, 1971; Kim et al, 1993).

Hawking radiation has not been observed after over two decades of searching (Halzen et al, 1991). Scientific papers (De Sabbata and Sivaram, 1992; Balbinot, 1986) have been written offering reasons why it may not be observable. For example, De Sabbata and Sivaram suggest that "Thus one may observe the decay [Hawking radiation] only if one makes an infinite succession of measurements. So in a sense one may never be able to observe the Hawking effect." The radiation described in the present paper differs substantially from Hawking's, and a case is made here that it has already been observed indirectly in ball lightning; and possibly also in the detected gamma-ray background.

## 2. Gravitational Tunneling

A quantum theory of gravity has not yet been developed. Nor has the difficult two-body problem yet been solved in general relativity. There may be an intrinsic incompatibility between general relativity and quantum mechanics as quantum mechanics appears to be antithetical to the equivalence principle (Overhauser, 1975; Rabinowitz, 1990a). Hawking (1974, 1975) boldly circumvented these problems in considering quantum fluctuation virtual particle pair production outside an isolated black hole as the source of Hawking radiation.

The tunneling probability is 0 out of the gravitational well of a single isolated body. Two-body systems will also be analyzed, where the tunneling probability $\neq 0$. The analysis here mainly relates to uncharged, non-rotating bodies in general, and black holes in particular. Since we shall be dealing mainly with very low tunneling probabilities, the details of the effective potential barrier near the black hole are not critical. Both for Einsteinian and Newtonian black holes, the potential energy far from the hole $\propto 1/r$. To avoid concerns related to Einsteinian black holes, we may consider that we are dealing with Newtonian black holes. Einstein himself was troubled with the nature of black holes in General Relativity. At present there is no direct or indirect experimental evidence concerning the space near or inside black holes, nor whether they are Einsteinian or Newtonian. The theory presented here provides predictions for indirect testing of the nature of black holes. The approach is similar in spirit to the prevalent approach of using a potential well to represent a nucleus although it is impossible to describe by a potential the forces acting on a particle inside the nucleus.

We could carry through a general abstract solution $e^{-2\Delta\gamma}$ in what follows. Since the difference between the Einsteinian and Newtonian gravitational potentials can be small, let us calculate specific transmission probabilities using the Newtonian potential. (For an isolated Einsteinian black hole, depending on angular momentum, there can be a barrier peaked at ~ 1.5 Schwarzchild radii = 1.5 $R_H$.)

**2.1 Isolated Body**

Even though the tunneling probability from the gravitational well of an isolated body is 0, let us derive it not only because gravitational tunneling appears not to have been done previously, but because the solution can give us an insight for the analysis of tunneling in the case of a gravitational potential due to more than one body, where the probability may be > 0.

The one-dimensional Schrödinger equation for a mass m in a well of potential energy V due to a spherical body of mass M centered at the origin is

$$\frac{-\hbar^2}{2m}\frac{d^2\Psi}{dr^2} + \Psi[V-E] = \frac{-\hbar^2}{2m}\frac{d^2\Psi}{dr^2} + \Psi\left[\frac{-GmM}{r} - E\right] = 0, \quad (1)$$

where the reduced mass $mM/(m+M) \approx m$, since $m \ll M$. (The essence of the one-dimensional approach survives generalization to three dimensions when there is spherical symmetry. The Hamiltonian for the attractive gravitational potential is of the same form as that of the hydrogen atom with radial wave function solutions in terms of Laguerre polynomials. However, we shall find it convenient to use the WKB approximation in what follows since the next model involves two bodies, and is roughly one-dimensional.)

As shown in Figure 1, the gravitational potential energy of a single isolated spherical body is $-GmM/r$ down to its surface, with total energy $E = -GmM/b_1$ at the classical turning point $b_1$. For a uniform mass distribution, the potential $\propto r^2$ inside the body. Since we are mainly interested in high energy solutions near the top of the well, we can neglect the bottom of the well, whatever its configuration. A wave function $\Psi$ of the form $\Psi = Ae^{-\gamma(r)}$ is a solution of eq. (1), when $d^2\gamma/dr^2 \approx 0$ is negligible.

The tunneling probability between points $b_1$ and $b_2$ is the ratio of probability densities at $b_1$ and $b_2$:

$$\Pi = \frac{\Psi(b_2)\Psi^*(b_2)}{\Psi(b_1)\Psi^*(b_1)} \approx e^{-2[\gamma(b_2)-\gamma(b_1)]} \equiv e^{-2\Delta\gamma} \qquad (2)$$

The solution for $\Delta\gamma \equiv \int_{b_1}^{b_2} \left[\frac{2m}{\hbar^2}(V-E)\right]^{1/2} dr$ that satisfies eq. (1) is

$$\Delta\gamma \approx \frac{m}{\hbar}\sqrt{2GM}\left\{\sqrt{\frac{b_2(b_2-b_1)}{b_1}} - \sqrt{b_1}\ln\left[\frac{\sqrt{b_2}+\sqrt{b_2-b_1}}{\sqrt{b_1}}\right]\right\}. \qquad (3)$$

Thus as expected there is no tunneling in this case as $\Pi = 0$, since as $b_2 \to \infty$, $\Delta\gamma \to \infty$.

**2.2 Black Hole Opposite Another Body**

A second body accomplishes two things. It lowers the barrier and gives the barrier a finite rather than infinite width, so that a particle can escape by tunneling or over the top of the lowered barrier. Only tunneling will be analyzed here. Black hole emission is greatest when the companion is a nearby almost black hole, and least when it is a distant ordinary body. The escaping particle may be trapped in the well of the second body. If it is not also a black hole, then escape from it can occur by ordinary excitation processes such as scattering and gravity-assisted energy from the second body's angular momentum.

From the symmetry of quantum tunneling for a non-absorbing barrier, the transmission amplitude and phase are the same in both directions (Cohn and Rabinowitz, 1990). This has significance for black hole tunneling in that the transmission probability must be the same into or out of a black hole. As derived in Section 2.1, the tunneling probability is 0 for escape from a single isolated black hole. Let us see quantitatively what effect a second body has on the tunneling probability.

As shown in Fig. 2, $M_2$ is centered at R opposite a black hole of mass M centered at the origin. Outside the black hole, we need to solve the Schrödinger equation

$$\frac{-\hbar^2}{2m}\frac{d^2\Psi}{dr^2} + \Psi\left[\frac{-GmM}{r} + \frac{-GmM_2}{R-r} - E\right] = 0 \qquad (4)$$

in the region $b_1 \leq r \leq b_2$, where $b_1$ and $b_2$ are the classical turning points, and $E = -GmM/b_1 + -GmM_2/(R-b_1) = -GmM/b_2 + -GmM_2/(R-b_2)$.

We solve for $\Delta\gamma$ as before:

$$\Delta\gamma \approx \frac{m}{\hbar}\sqrt{\frac{2GM}{d}}\left\{\sqrt{b_2(b_2-d)} - \sqrt{b_1(b_1-d)} - d\ln\left[\frac{\sqrt{b_2}+\sqrt{b_2-d}}{\sqrt{b_1}+\sqrt{b_1-d}}\right]\right\} \qquad (5)$$

where $d = Mb_1(R-b_1)R/[M(R-b_1)R+M_2(b_1)^2]$, and the solution applies for $R \gg b_2$, when $M_2 \gg M$. Eq. (5) reduces to Eq. (3) for $R \to \infty$, as it should, and $\Pi \to 0$. As in the previous case $\Delta\gamma \to 0$ for $b_1 \to b_2$, yielding $\Pi \to 1$. When $M \to 0$, or $M_2 \to \infty$, or equivalently $[M/M_2] \to 0$, $\Delta\gamma \to 0$ and $\Pi \to 1$. Eq. (5) can serve as a lower limit check on $\Pi$ when exact calculations can't be done for the general two-body case.

The mass $M_2$ can be fixed in space, orbit around the black hole, or be at R temporarily. A mutually orbiting black hole and mass $M_2$ will produce a lighthouse effect for an observer who can detect well-timed gamma-ray pulses when the black hole, orbiting mass, and observer line up. Radiation will escape from the black hole as long as $M_2$'s lingering time >> tunneling time or black hole transit time, whichever is greater. Although in the last decade there has been a dramatic increase in both our experimental and theoretical knowledge of gamma-ray pulsars (Yadigaroglu and Romani, 1997) thanks to the data provided by the Compton Gamma Ray Observatory, there still remain somewhat unanswered questions to which such holes may shed light.

The tunneling calculations in Section 2 are general and also apply to gravitational tunneling of ordinary bodies. It is remarkable that a black hole of infinite mass in the presence of another body becomes completely transparent quantum mechanically ($\Pi = 1$). Nevertheless as we shall see, it cannot radiate, since the radiated power $\propto [1/M]^2 \to 0$ as $M \to \infty$. As an isolated body, any black hole would be completely opaque ($\Pi = 0$); but there are always other bodies, e.g. the rest of the universe.

### 3. Transmission Probability

A distinction must be made between the concepts of "transmission probability or transmission coefficient" and "tunneling probability or penetration coefficient." The first is a ratio of probability densities and the second is a ratio of probability current densities. The much earlier literature often did not distinguish between the two concepts and this still occurs occasionally. The transmission probability or coefficient

$$\Gamma \equiv \frac{\Psi(b_2)\Psi^*(b_2)v_2}{\Psi(b_1)\Psi^*(b_1)v_1} \approx \frac{v_2}{v_1} e^{-2[\gamma(b_2)-\gamma(b_1)]} = \Pi \frac{v_2}{v_1} , \qquad (6)$$

where tunneling is from region 1 (left of the barrier) to region 2 (right of the barrier). $\Gamma = \Pi$ when the velocities $v_1$ and $v_2$ are the same on both sides of the barrier.

Let us see in general for any barrier approximately how $\Gamma$ and $\Pi$ are related to the energy E of the particle and an arbitrary potential barrier V. Following similar analysis to that in Section 2, in region 2

$$\Psi_2 \approx \left[\frac{2m}{\hbar^2}(E-V)\right]^{-\frac{1}{4}} \exp(i)\left(\int_{b_2}^{r}\left[\frac{2m}{\hbar^2}(E-V)\right]^{\frac{1}{2}}dr\right), b_2 \leq r . \qquad (7)$$

Matching the magnitude and first derivative of the wave function $\Psi_3$ in the classically forbidden region 3 inside the barrier with $\Psi_1$ and $\Psi_2$ at $b_1$ and $b_2$,

$$\Psi_3 \approx \left[\frac{2m}{\hbar^2}(E-V)\right]^{-\frac{1}{4}} e^{-i\frac{\pi}{4}}\left(e^{B_3} + \tfrac{1}{2}ie^{-B_3}\right), b_1 \leq r \leq b_2, \text{ where} \qquad (8)$$

$$B_3 = \int_r^{b_2}\left[\frac{2m}{\hbar^2}(V-E)\right]^{\frac{1}{2}} dr$$

$$= \int_{b_1}^{b_2}\left[\frac{2m}{\hbar^2}(V-E)\right]^{\frac{1}{2}} dr - \int_{b_1}^{r}\left[\frac{2m}{\hbar^2}(V-E)\right]^{\frac{1}{2}} dr \equiv \Delta\gamma - B_1 \qquad (9)$$

Substituting eq. (9) into eq. (8)

$$\Psi_3 \approx \left[\frac{2m}{\hbar^2}(E-V)\right]^{-\frac{1}{4}} e^{-i\frac{\pi}{4}}\left(e^{\Delta\gamma - B_1} + \tfrac{1}{2}ie^{-\Delta\gamma + B_1}\right). \text{ Thus} \qquad (10)$$

$$\Psi_1 = \left[\frac{2m}{\hbar^2}(E-V)\right]^{-\frac{1}{4}} \left\{\begin{array}{l} -i\left[\exp i\int_r^{b_1}\left[\frac{2m}{\hbar^2}(E-V)\right]^{\frac{1}{2}} dr\right]\left[\left(e^{\Delta\gamma}-\tfrac{1}{4}e^{-\Delta\gamma}\right)\right] + \\ \left[\exp-i\int_r^{b_1}\left[\frac{2m}{\hbar^2}(E-V)\right]^{\frac{1}{2}} dr\right]\left[\left(e^{\Delta\gamma}+\tfrac{1}{4}e^{-\Delta\gamma}\right)\right] \end{array}\right\} = \Psi_{inc} + \Psi_{ref}. \quad (11)$$

Now $\Psi_1$ has been expressed as a sum of an incident and a reflected wave, where the incident wave is

$$\Psi_{inc} = -i\left[\frac{2m}{\hbar^2}(E-V)\right]^{-\frac{1}{4}} \left[\exp i\int_r^{b_1}\left[\frac{2m}{\hbar^2}(E-V)\right]^{\frac{1}{2}} dr\right]\left[\left(e^{\Delta\gamma}-\tfrac{1}{4}e^{-\Delta\gamma}\right)\right], \qquad (12)$$

In this general case, without needing an explicit solution for $\Psi$,

$$\Gamma = \frac{\Psi_2 \Psi_2^*}{\Psi_{inc}\Psi_{inc}^*}\left(\frac{v_2}{v_1}\right) = \left[e^{\Delta\gamma} - \tfrac{1}{4}e^{-\Delta\gamma}\right]^{-2} \qquad (13)$$

From eq. (9),

$$\Delta\gamma \equiv \int_{b_1}^{b_2}\left[\frac{2m}{\hbar^2}(V-E)\right]^{\frac{1}{2}} dr. \qquad (14)$$

Thus when $\Delta\gamma$ is large, $e^{\Delta\gamma} \gg (1/4)e^{-\Delta\gamma}$ in eq. (13), yielding $\Gamma \approx \Pi = e^{-2\Delta\gamma}$. $\Gamma \approx \Pi$ is true in most cases when $b_2 \gg b_1$, and/or $V \gg E$. Note that $e^{-2\Delta\gamma}$ is the solution obtained for the two cases in Section 2, where $\Delta\gamma$ was obtained via the integral of eq. (14).

However we shall be mainly interested in the high energy case, when V - E is small, which (for our gravitational barriers) implies that the distance between the classical turning points, $b_2 - b_1$, may not be relatively large. At first sight it would appear that we cannot make the approximation $\Gamma \approx \Pi$. Propitiously, the barrier of Section 2.2 becomes symmetrical for all energies and barrier widths when $M = M_2$, and then $v_1 = v_2$. Similarly $v_1 \sim v_2$ for $M \sim M_2$. So in this paper $\Gamma \approx \Pi$ is a valid approximation when $M \sim M_2$ and is true for all $M$ and $M_2$ in the case of ultrarelativistic electrons and positrons, photons ($m = h\nu/c^2$), and neutrinos where $v_1 \approx v_2 \approx c$, the speed of light. However, for non-zero rest mass particles, when their energies are low in a non-symmetrical gravitational barrier, this may not be a valid approximation. This seems to have been neglected by Hawking and others. It is a good approximation for for little black holes because low energy particles are a miniscule fraction of the radiation due to the extremely high temperature, but need to be taken into consideration for intermediate and high mass black holes.

### 4. Emission Rate

Complementary procedures may be used in calculating the emission rate from a black hole. In one, the probability current density or flux

$$j = \frac{\hbar}{2im}\left[\Psi^* \mathrm{grad}\Psi - \Psi \mathrm{grad}\Psi^*\right] \tag{15}$$

is integrated as a dot product over the surface area of the black hole (or ordinary body) to yield the emission rate. However, it is misleading to consider j to be the particle flux or average particle flux at <u>a given point</u>. For a precise measurement of even the average local flux implies simultaneous high-precision measurements of position and velocity (equivalent to momentum) which would lead to a violation of the uncertainty principle. However, it is heuristically useful to treat **j** as a flux vector, especially when it has weak or no dependence on position, allowing an accurate velocity determination. Mashhoon (1990) analyzes other limitations which can be helpful in considering the multi-faceted problems related to black hole radiation.

A procedure is taken here similar to that traditionally used for tunneling out of a nucleus. Each approach of the trapped particle to the barrier has the calculated probability of escaping or tunneling through the barrier. Thus we need only know the frequency of approach to the barrier. Heuristically, the time between successive impacts on the barrier for ultrarelativistic particles is

$$\tau = \frac{2\langle r \rangle}{c} \approx \frac{2R_H}{c} = \frac{2(2GM/c^2)}{c} \, , \tag{16}$$

where $R_H$ is the Schwarzchild radius. The Schwarzchild radius determines a spherical surface of classical no return for a non-rotating black hole.

Thus in the high energy case, the emission rate or probability of emission per unit time from the black hole is

$$\frac{\Gamma}{\tau} = \frac{\Pi}{\tau}\frac{v_2}{v_1} = \frac{\Pi}{\tau}\frac{c}{c} = \frac{\Pi}{\tau} \approx \frac{\Pi c^3}{4GM} = \frac{e^{-2\Delta\gamma}c^3}{4GM} \, , \tag{17}$$

where $\Delta\gamma$ is given by eqs. (3) and (5) for the models discussed in Section 2. Eq. (17) is a good approximation when a little black hole is opposite a large body such as the earth. A fractional solid angle, $\Delta\Theta/4\pi$, reduced emission correction factor needs to multiply eq. (17) when the adjacent body is small.

### 5. Black Hole Radiated Power

The mean power radiated from a black hole of volume $\Omega$ is

$$P_R = \frac{\int \Psi^* E_e \frac{\Gamma}{\tau} \Psi d\Omega}{\int \Psi^* \Psi d\Omega} \approx \left\langle E_e \frac{\Pi}{\tau} \right\rangle \sim \langle E_e \rangle \frac{\langle e^{-2\Delta\gamma} \rangle c^3}{4GM} \, , \tag{18}$$

where $\langle E_e \rangle \approx kT$ is the average energy of the emitted photon, and

$$\langle E_e \rangle = \frac{\int \Psi^* E_e \Psi d\Omega}{\int \Psi^* \Psi d\Omega} \, . \tag{19}$$

Although $e^{-2\Delta\gamma}$ is the same at the same energy from the second body into the black hole, energy degradation in its well greatly reduces the tunneling rate back into the hole.

The Hawking expression for temperature is derived on the basis of entropy considerations (Bekenstein, 1973, 1974). Hawking's 1974 value is a factor of 2 smaller than his 1975 value. This is not critical, and the 1975 expression is

$$T = \left[\frac{\hbar c^3}{4\pi kG}\right]\frac{1}{M} = [2.46 \times 10^{23}]\left(\frac{1}{M}\right){}^\circ K \, , \tag{20}$$

with M in kg.  For M ~ $10^{12}$ kg (the largest mass that can survive to the present for Hawking), T ~ $10^{11}$ K.  As we shall see, the new theory permits the survival of much smaller masses, such as for example M ~ $10^6$ kg with T ~ $10^{17}$ K.

Combining eq. (20) with eq. (18) for the tunneling radiation power:

$$P_R \approx \left[\frac{\hbar c^3}{4\pi GM}\right]\frac{\langle e^{-2\Delta\gamma}\rangle c^3}{4GM} = \left[\frac{\hbar c^6 \langle e^{-2\Delta\gamma}\rangle}{16\pi G^2}\right]\frac{1}{M^2} = \frac{\langle e^{-2\Delta\gamma}\rangle}{M^2}\left[3.42\times10^{35}\,\text{W}\right]. \qquad (21)$$

Note that $P_R$ was obtained without invoking field fluctuations, pair creation, quantum fluctuations of the metric, etc.  No correction for gravitational red shift needs to made since the particles tunnel through the barrier without change in energy.  Using the Hawking temperature may appear inconsistent.  It is used herein since an unshifted dynamical black hole temperature can be derived which is close to that of eq. (20).  Although originally proposed as not being real, this temperature is now asserted and generally accepted as being the gravitationally red shifted temperature (Bardeen, Carter, and Hawking, 1973). Their new view implies an infinite temperature at the horizon of all black holes, since this red shift goes to 0 as measured at large distances from any hole if the surface temperature were finite.  For the real temperature, they said "the effective temperature of a black hole is zero ... because the time dilation factor [red shift] tends to zero on the horizon."  This question deserves further consideration.

The Hawking radiation power, $P_{SH}$, follows the Stefan-Boltzmann radiation power density law $\sigma T^4$, when $\frac{8\pi GME}{\hbar c^3} \gg 1$.  For Hawking :

$$P_{SH} \approx 4\pi R_H^2 [\sigma T^4] = 4\pi\left(\frac{2GM}{c^2}\right)^2 \sigma\left[\frac{\hbar c^3}{4\pi kGM}\right]^4 = \frac{\hbar^4 c^8}{16\pi^3 k^4 G^2}\{\sigma\}\left[\frac{1}{M^2}\right], \qquad (22)$$

where $\sigma$ is the Stefan-Boltzmann constant.  Although $P_R$ and $P_{SH}$ appear quite disparate, the differences almost disappear if we substitute into eq. (22) the value obtained for $\sigma$ by integrating the Planck distribution over all frequencies:

$$\sigma = \left\{\frac{\pi^2 k^4}{60\hbar^3 c^2}\right\}, \qquad (23)$$

$$P_{SH} = \frac{\hbar^4 c^8}{16\pi^3 k^4 G^2}\left\{\frac{\pi^2 k^4}{60\hbar^3 c^2}\right\}\left[\frac{1}{M^2}\right] = \frac{\hbar c^6}{16\pi G^2}\left\{\frac{1}{60}\right\}\left[\frac{1}{M^2}\right]. \tag{24}$$

Thus
$$P_R = 60\langle e^{-2\Delta\gamma}\rangle P_{SH}. \tag{25}$$

It is remarkable that even though $P_R \propto T$ and $P_{SH} \propto T^4$, they can be put into an equivalent form, aside from the numerical factor $60\langle e^{-2\Delta\gamma}\rangle$.

## 6. Black Hole Evaporation

The evaporation rate for a black hole of mass M is $d(Mc^2)/dt = -P_R$, which gives the lifetime

$$t = \frac{16\pi G^2}{3\hbar c^4 \langle e^{-2\Delta\gamma}\rangle}[M^3]. \tag{26}$$

This implies that the smallest mass that can survive up to a time t is

$$M_{small} = \left(\frac{3\hbar c^4 \langle e^{-2\Delta\gamma}\rangle}{16\pi G^2}\right)^{1/3}[t^{1/3}]. \tag{27}$$

Primordial black holes with $M \gg M_{small}$ have not lost an appreciable fraction of their mass up to the present. Those with $M \ll M_{small}$ would have evaporated away long ago.

Thus the smallest mass that can survive within $\sim 10^{17}$ sec (the age of our universe) is

$$\mathbf{M_{small} \geq 10^{12}\langle e^{-2\Delta\gamma}\rangle^{1/3}\text{ kg}}. \tag{28}$$

Inasmuch as $0 \leq e^{-2\Delta\gamma} \leq 1$, an entire range of black hole masses much smaller than $10^{12}$ kg may have survived from the beginning of the universe to the present than permitted by Hawking's theory. For example, if the average tunneling probability $\langle e^{-2\Delta\gamma}\rangle \sim 10^{-18}$, then $M_{small} \sim 10^6$ kg, and these bodies will presently radiate at $10^6$ higher temperature; and $\sim [10^{12}/10^6]^2 = 10^{12}$ times more power than a $10^{12}$ kg black hole with the same $e^{-2\Delta\gamma}$.

These differences in the expected radiation may help to explain why the Hawking radiation profile (Halzen et al, 1991) has not yet been detected. It seems that the present theory can be helpful in understanding the observed gamma-ray background, which has far more photons at higher photon energy than expected from the Hawking model.

### 7. Dark Matter

We do not know what 95 % of the universe is made of. One piece of evidence that there must be 95% dark matter or missing mass comes from spiral galaxies. There must be some unseen form of matter whose gravitational attraction is great enough to hold the galaxies together as they rotate, as discovered by the unheralded Vera Rubin (1983). The missing mass gives the stars ~ constant linear velocities independent of radial distance r, rather than the expected Keplerian velocities $\propto 1/\sqrt{r}$. The rate of rotation is so great that they would fly apart if they contained only the stars and gas we can directly perceive. Another piece of evidence for dark matter comes from clusters of galaxies. Galaxies are gathered together in clusters that range from a few galaxies to millions. These clusters exist because the galaxies attract each other into groups. The speeds at which individual galaxies are moving in these clusters are so high that the clusters would fly apart unless they were held together by a stronger gravitational attraction than provided by the masses of all the galaxies.

It is possible that the early universe underwent a phase transition. In a phase transition an initially uniform medium develops irregularities -- in the case of freezing and boiling of water these are clumps of ice or bubbles of steam. Hawking (1971) proposed that these irregularities collapsed to form little primordial black holes. Such black holes could have been present in the early universe at the time of nucleosynthesis, and may have affected its results if the radiation emitted by them either interfered with the nucleosynthesis itself or broke up products of nucleosynthesis after the nuclear reactions were over. The little quiescent black holes derived herein are much less likely to interfere with nucleosynthesis than Hawking's, and thus can be much smaller than previously considered. Thus primordial black holes would not be subject to the limits imposed by nucleosynthesis arguments. The baryonic matter that got trapped and crushed in them would have bypassed the deuterium and helium formation that occurred during the era of nucleosynthesis.

Small black holes as envisaged here are quiescent compared with Hawking's. Since they are so extremely massive for their miniscule size, they may well explain the missing mass or so-called dark matter of which ~ 95% of the universe is composed. Because they can be small compared with the wavelength of visible light, they will not scatter or occlude light from the distant stars. For example, black holes of between $10^{-7}$ kg and $10^{19}$ kg have radii between $10^{-30}$ m and $10^{-8}$ m = 100 Å, well below visible wavelengths. To account for the missing dark matter there would need to be between $10^{61}$ and $10^{35}$ such black holes for a universe mass of ~ $10^{53}$ kg (Rabinowitz, 1990b, 1998). For our universe of radius $15 \times 10^9$ light-year = $1.4 \times 10^{26}$ m, this would require an average density of between $10^{30}$ and $10^5$ black holes per cubic light-year, and more than this near galaxies. This is orders of magnitude larger than permitted for Hawking's extremely radiative black holes. That many of his little black holes would fry the universe.

Assuming an initially approximately uniform distribution of little black hole mass, stars in young galaxies will orbit with ~ constant angular velocity, i.e. ~ constant period. As the age of the universe gets extremely long with respect to the present age of ~ $10^{17}$ sec., the stars will orbit with ~ Keplerian periods $\propto r^{3/2}$. In the present epoch, stars orbit with ~ constant linear velocity because the little black hole total mass increases with radial distance from a galactic center due to radiation reaction force driving them outward, as well as a lower evaporation rate at larger radial distances. This mechanism may also be able to account for the recently observed accelerated expansion of the universe.

Even with the dark matter being 95% of the mass of the universe, on the small solar system size scale the sun's mass dominates over any kind of missing mass since the volume of the solar system is not large enough to hold enough missing mass based on its average universal density. Thus there is no appreciable deviation from Keplerian motion of the planetary orbits. The distribution of black hole masses in the universe has not been determined experimentally or even estimated theoretically. Gravitational fields of other bodies can both enhance the number density of little black holes, and act locally as a mass filter by radiation reaction force repulsion.

## 8. Ball Lightning Analysis

Ball lightning is one of the few long-known and widely-accepted natural phenomena which are still unexplained. Even though Trofimenko (1990) presented a large list of potential astrophysical and geophysical phenomena that might be affected by little black holes, ball lightning was not among them. It is easy to see why the scientific community has not considered little black holes as the core power source of ball lightning because Hawking's little black holes radiate at a devastatingly high rate in all directions (1974, 1975) that would hardly go unnoticed. Prior to the awareness that black holes can radiate, their presence in the earth would have been considered highly unlikely as the earth would have been devoured after ~ $10^6$ years, leaving a black hole of 1 cm radius. But the earth has existed for over $4 \times 10^9$ years. For the new view of little black holes, the downward directed radiation between the hole and the earth can provide levitation, with a small horizontal component providing mobility, and the holes radiate considerably less. When they get so small that there would be appreciable radiation, the radially outward radiation reaction force propels them away from the earth. For these and many other reasons that we shall see, they are excellent candidates as the source of ball lightning.

Let us make some estimates that are illustrative rather than strictly quantitative. From eqs. (5) and (28) with the earth as the second body, $M_2 = M_{earth} = 6 \times 10^{24}$ kg ,

R = 6 x $10^{26}$ m, and a mean transmission coefficient $\langle\Gamma\rangle \sim 10^{-37}$, little black hole masses as small as 1/2 kg can survive from the early universe in the region of the earth. Even smaller masses down to $10^{-7}$ kg that are remnants of larger masses can be present, as well as miniscule primordial little black holes from outer space that previously had $\langle\Gamma\rangle << 10^{-37}$.

The downwardly directed radiation (due to the earth below) from a 3 x $10^{-4}$ kg ($\approx$ 1/3 gm) little black hole will act like a rocket exhaust permitting the little black hole to levitate or fall slowly. We can estimate the upward force on the little black hole from

$$M\frac{dv}{dt} = -c\frac{dM}{dt} - Mg \qquad (29)$$

where the exhaust leaves the little black hole at the speed of light, c= 3 x $10^8$ m/sec, and g = 9.8 m/sec$^2$ is the acceleration of gravity near the earth's surface. For the above values, and eq. (21) for $dM/dt = -P_R/c^2$, eq. (29) shows that with negligible initial downward velocity the little black hole will fall slowly from a height of 3m with an acceleration of $\sim 10^{-1}$ m/sec$^2$, disappearing into the ground in $\sim$ 10 sec. Similar disappearance times would be obtained for entry into nearby structures or stasis.

From Eqs. (20) and (21), a 1/3 gm little black hole has a temperature T $\sim 10^{27}$ K, and radiates $\sim 10^6$ W ($\sim$ 37% by electrons and positrons, $\sim$ 8 % by photons, and the remainder equally divided by the six kinds of neutrinos together with a very small component of gravitons). From the Planck black body radiation distribution we can calculate the fraction of the power in the visible spectrum. The total power per unit area emitted by a black body is

$$\mathbf{P/A} = 2\pi hc^2 \int_0^\infty \frac{d\lambda}{\left(e^{ch/\lambda kT}-1\right)\lambda^5} = \left\{\frac{\pi^2 k^4}{60\hbar^3 c^2}\right\} \mathbf{T^4}, \qquad (30)$$

which is the Stefan-Boltzmann law. For the visible part of the spectrum

$$(\frac{P}{A})_{vis} = 2\pi hc^2 \int_{\lambda_1}^{\lambda_2} \frac{d\lambda}{[(e^{ch/\lambda kT})-1]\lambda^5} \approx \int_{\lambda_1}^{\lambda_2} \frac{d\lambda}{[(1+ch/\lambda kt)-1]\lambda^5} = \frac{2\pi c^2 h}{3}\left(\frac{kT}{ch}\right)\left[\frac{1}{\lambda_1^3} - \frac{1}{\lambda_2^3}\right] (31)$$

where the exponential has been expanded to first order. The ratio of equation (31) to (30) is $\sim 10^{-67}$, where the wavelength range for the visible spectrum is taken to be $4000$ Å $\leq \lambda \leq 8000$ Å. Since only ~ 8% of the little black hole radiation is in photons, only ~ $10^{-68}$ [$P_R$] = $10^{-68}$ [$10^6$ W] = $10^{-62}$ W, of this power is in the visible. Thus a powerfully radiating little black hole is not directly visible.

Little black holes become visible indirectly as ball lightning in the surrounding air by excitation and direct collisional ionization with a charged little black hole resulting in electron ion pair recombinations, by excitation of the air molecules and atoms from the $4 \times 10^5$ W of power emission of electrons and positrons, and by infalling particle collisions. Let us estimate the efficiency of this process. In terms of ionization, the number of electron-ion pairs that can be produced by the black hole's local deposition of energy, $E_{local}$ is

$$N = \eta \frac{E_{local}}{eV_i} , \qquad (32)$$

where e is the electronic charge, $\eta$ is the ionization efficiency, and the average ionization potential is $V_i = 0.8\ (15.5\ V)_{N_2} + 0.2\ (12.5\ V)_{O_2} = 14.9$ V. If neutral, the little black hole in moving through air will lose only ~ $10^{-21}$ ergs/cm to the surrounding atmosphere (Greenstein and Burns, 1984). It is interesting to note that if it were not for accelerations due to the tunneling emission, neutral little black holes would have close to the original velocity with which they were created because of their tiny interaction cross section. If charged internally with ~ 10 electron charges, the energy loss could be as high as ~ $10^{11}$ ergs/cm, and even higher if charged externally as discussed at the end of this section. This energy loss increases proportionately to the density of the medium, and as the square of the total charge. If the ionization efficiency $\eta$ were unity, this would produce as many as $10^{28}$ electron-ion pairs/cm of path. Estimating a velocity > $10^2$ cm/sec with a visible lifetime ~ 10 sec, there would be enough ionization for both a visible and a radar signal. However most of the local energy deposition goes into heating up the air with elastic and inelastic collisions that are not ionizing. So we need to estimate the ionization efficiency.

The maximum in the ionization cross-section occurs at ~ 10 $V_i$. To get an insight into the ionization efficiency η, let us take for example electrons colliding with atmospheric molecules at the rate of ~ $10^{10}$/sec to $10^{12}$/sec in a uniform electric field, Σ, producing ionization:

$$\eta = \frac{V_i \alpha}{\Sigma}, \tag{33}$$

where α is the first Townsend ionization coefficient. For air at atmospheric pressure, the low frequency breakdown field is Σ ≈ 3000 kV/m and α is 900/m. This implies that for an ordinary air discharge the ionization efficiency η is only 0.5%. Since the vast majority of emitted particles have energy >>> 10 $V_i$, η is many orders of magnitude lower than this for the emitted particles. A slow charged little black hole would be much more efficient at producing ionization and excitation than the emitted particles.

Because almost all the emitted particles are ultrarelativistic, there will also be a small Cerenkov radiation contribution to the emitted energy. For a charged little black hole moving very close to or into a conductor, there will also be Lilienfeld (1919) transition radiation appearing to come out of the conductor due to the time-varying virtual dipole between the charge and the image charge in the conductor. It has an easily identified signature as transition radiation is plane polarized. A small amount of bremsstrahlung radiation is expected as the little black hole enters ordinary low atomic number (Z) walls and other structures, because there will only be a small deceleration, which would increase as Z goes up.

An upper limit order of magnitude overall excitation efficiency for all processes is f < $10^{-5}$. The visible power of a cool glowing ball of air of radius ~$10^{-1}$ to 10 cm is

$$P_{vis} = f [P_R] \sim 10^{-5}[(10^{11} \text{ erg/cm})(10^2 \text{ cm})/\text{sec}] \sim 10 \text{ W} \tag{34}$$

surrounding a charged little black hole core moving through the atmosphere. A little black hole at the center of glowing ball lightning has similarities on a relatively miniscule scale to a galactic black hole and accretion disk. The huge disk radiates with enormous protruding jets because of the high speed and high collision rate of molecules falling into the extremely massive but relatively much smaller (in dimension) galactic black hole.

A little black hole can trap charge internally and/or externally. It could easily trap ~ 10 positive or negative charges externally and form a neutral or charged super-heavy atom-like structure. The circulating negative or positive charge that survives collision with the high energy radiation would tend to move in a plane perpendicular to the emission until precession brings it into a collision with this radiation, eventually causing extinction of the ball lightning. The mass input due to infalling matter is more than countered by a decrease in mass due to the radiation emission. Accumulation of matter into the little black hole of ~ 1 gm mass would be somewhat limited to particles with a de Broglie wavelength < the Schwarzchild radius, $R_H \sim 5 \times 10^{-28}$ cm. Thus neutralization of the internal charge of a little black hole would not occur as rapidly as one might otherwise expect and luminous lifetimes ~10 to 1000 sec may be achievable.

### 9. Meeting Ball Lightning /Earth Lights Criteria

If greatly decreased radiation permits little black holes to be prevalent throughout the universe, then it is reasonable to surmise that they are also present in the region of the earth. If they are present on earth, then one may ask how they might manifest themselves. If their presence can help to explain a long-known, well-established phenomenon that has no other explanation, then they are viable candidates for experimental investigations to test the validity of this hypothesis. It appears that ball lightning/earth lights represent an admirable testing ground.

A subtle variety of ball lightning are atmospheric luminous phenomena occurring in locations such as Hessdalen, Norway and elsewhere in the world. These are sometimes called "earth lights" (Devereaux, 1989), to make a refined distinction between them and ball lightning, as they appear to be more dynamic and unrelated to thunderstorm activity though otherwise they are very similar. This may just be a manifestation of little black holes where there is a large component of horizontal velocity due to a small component of horizontal radiation reaction force due to the presence of mountains at Hessdalen..

At Hessdalen large numbers of researchers have observed earth lights moving parallel to the earth. The sightings were visual, photographic, and had strong radar signals (Strand, 1984). Such observations are compatible with a charged levitating black hole. The luminosity and radar signals may be accounted for by the atmospheric ionization created by a charged little black hole and dragged along by electrostatic attraction to the hole. Lifetime measurements of the (ball lightning-like) earth lights at Hessdalen are among the most reliable as these were directly measured by numerous well-prepared observers both optically and with radar .

The following criteria are presented as a guide for assessing ball lightning/earth light models in general, and the little black hole model in particular. The first five are derived from Uman (1968), and the rest are inferred from several sources (Bach, 1993; Fryberger, 1994; and Singer, 1971).

**1) Constant size, brightness, and shape for extended times**

The large amount of gravitationally stored energy in little black holes and resulting kinetic energy accounts for the somewhat constant size, shape, and brightness of ball lightning; and its particular shape is a function of the motion of the little black hole as it drags along ionized air. Ball lightning has stable spherical, pear-shaped, prolate and oblate ellipsoidal, cylindrical, and disk shapes (Bach, 1993; Singer, 1971).

Models that depend on thermally stored energy do not have stability due to cooling with time. As given by eq. (26) the little black hole used above for example estimates can have a lifetime ~ 1 year near the earth. As it evaporates to a much smaller mass, with a concomitant increase in radiation reaction force, it will shoot up into space and thus extend its lifetime. Its luminosity can vanish when its trapped charge becomes neutralized, by going into the ground or other opaque structures, or when the black hole itself becomes disrupted, as possibly when the electrostatic repulsive force of the captured charge ≈ its gravitational force.

There are a number of models that fit this criterion. Finkelstein and Rubinstein (1964) proposed that ball lightning is a luminous region of air of nonlinear high electrical conductivity carrying a high current density. They showed that their model can yield ball-like solutions. A similar theory was presented by Uman and Helstrom (1966). Winterberg (1978) proposed an electrostatic theory of ball lightning.

**2) Untethered high mobility**

The lightness of the little black hole (~ 1/3 gm in the example calculation) in which the ball lightning mass mainly resides, gives it high mobility. A small horizonal component of the exhaust force accounts for its horizontal mobility. A charged black hole will also experience an attractive force towards its image charge in a conductor, and either a repulsive or attractive force with a charged dielectric, depending on the sign of the charge. Untethered mobility vitiates against electrical discharge models of ball lightning which require attachment to good (e.g., metal) or poor conductors (e.g., earth, wood) such as for St. Elmo's fire.

**3) Generally doesn't rise**

The ball lightning ionized air is electrostatically bound to the charge trapped in the little black hole and so is forced to follow its trajectory rather than simply rise. Since heated air expands and rises, this is another criterion against thermal source ball lightning. Occasionally, ball lightning ascends faster than possible for heated air. Masses << 1/3 gm would rapidly ascend and vanish from the atmosphere. The majority of ball lightning observations are of a slow descent.

**4) Can enter open or closed structures**

The radius of a $3 \times 10^{-4}$ kg (~1/3 gm) little black hole is $R_H = 2GM/c^2 = 5 \times 10^{-31}$ m. Uncharged little black holes have mean free paths through matter $>> 10^6$ km (Greenstein and Burns, 1984), and the mean free path of charged black holes $>>$ m. Little black holes thus can easily penetrate through any material. Ohtsuki and Ofuruton (1991) have created plasma fireballs formed by microwave interference in air containing ethane and/or methane. These fireballs evidently can penetrate dielectric materials, but not metals. They may have difficulty meeting the requirement of low optical power. Smirnov (1990) and others have presented strong arguments that ball lightning cannot be a plasmoid. This criterion militates against most models that require external energy sources.

### 5) Can exist within closed conducting metal structures

Since little black holes have a more than adequate supply of stored energy they can easily exist inside any closed highly conducting structure. However, this criterion dictates against models that depend on electrical currents, microwaves, or other electromagnetic radiation that is shielded out by a conductor. Microwave models such as that of Kapitza (1968), Ohtsuki and Ofuruton (1991) and others would be ruled out in this case.

### 6) Levitation

The little black hole's downwardly directed radiation accounts for steady levitation. It is hard for other models to account for steady levitation while moving horizontally for long distances without rising.

### 7) Low power in the visible spectrum

In the example, although the little black hole emits $10^6$ W of total power, it only produces < 10 W of optical power by ionization of the surrounding air. The bulk of ball lightning observations (Bach, 1993; Singer, 1971) suggest that the observed intensities of light and heat are $< \sim 10$ W . This criterion rules out all those models for which the total visible radiated power would be far too great for the appropriate color temperature of the ball lightning.

### 8) Rarity of sightings

Almost everyone has seen lightning, but few people have seen ball lightning. Since little black holes are quite rare, this explains the rarity of sightings. Many models are not in accord with this criterion. With galactic concentration of the 95% black hole dark matter, ~ $10^3$ little black holes may be expected in the region of the earth of volume $10^{12}$ km$^3$ (256 cubic billion miles) with ~ $1/10^9$ km$^3$ (~1 per cubic billion miles).

**9) Relatively larger activity near volcanoes**

Bach (1993) documents a relatively larger activity of ball lightning near volcanoes. Given Trofimenko's proposal (1990) that LBH are the main source of heat for volcanoes, it follows that little black hole-caused ball lightning should be more prevalent there. Other models don't explain this.

**10) Extinguishes quietly**

Ball lightning from little black holes extinguishes its luminosity quietly when it enters opaque materials like the ground or structures, slows down considerably, comes to rest, or becomes neutralized.

**11) Extinguishes explosively occasionally**

Ball lightning sometimes releases energy explosively (Bach, 1993; Singer, 1971). Little black holes occasionally extinguish explosively as their mass $\to 10^{-8}$ kg, or when otherwise disrupted. In 1846, lightning accompanied by fire-balls that "descended and exploded with terrific force" demolished the stone steeple of St. George's church in Leicester. In examining the remains of the steeple apex, Mills (1971) detected no radioactivity. He considered that radioactivity may have been undetectable because of the 125 years time lapse, but could be detectable "within days of a ball lightning strike."

**12) Related radioactivity**

Mills was testing the Altschuler et al (1970) model that ball lightning arises from a concentration of short-lived radioisotopes produced by lightning. There can be a low-level of γ-rays, positrons, and other radioactivity associated with ball lightning (Singer, 1993). Ashby and Whitehead (1971) tested the hypothesis that ball lightning is caused by antimatter meteorites. They made radiation measurements over the period of one year near thunderstorms and tornadoes to check whether the annihilation of minute fragments of meteoric antimatter in the upper atmosphere could be the cause of ball lightning. Though radioactivity was detected, they seem to have disproved both the Altschuler et al hypothesis and their own model. little black holes can account for radioactivity, whereas most other models cannot.

### 13) Typical absence of associated deleterious effects

Because of the low interaction cross section of the emitted radiation, the great total emitted power from a little black hole has low power density near the earth and low local power deposition. dissipating over a huge volume.

### 14) Occasional high localized energy deposition

Some ball lightning incidents require MJ of energy to account for molten materials and boiling away of a bathtub full of water (Bach, 1993; Singer, 1971). The high energy content >> MJ of little black holes can account for this when a little black hole is disrupted by an end of life burst; or moves through a much higher density material than air.

### 15) Larger Activity Associated With Thunderstorms

Thunderstorm activity may be involved in the charging of little black holes, and/or the high fields (Cobine, 1958) associated with thunder clouds may attract charged little black holes. During lightning, runaway high energy charged particles in the high energy tail of the Maxwellian distribution have more of a chance of being captured by the black hole due to their shorter de Broglie wavelengths. The potential of charged clouds may get as high as $10^9$ V (Rabinowitz, 1987).

## 10. Conclusion

The goal of this paper has been to present an alternative model which can be experimentally tested. Hawking radiation has proven elusive to detect. This paper gives an insight as to why this may be so. In a very young compact universe, radially directed tunneling radiation would have been substantial and may have contributed to the expansion of the early universe. In later epochs this radially directed radiation can help clarify the recently discovered accelerated expansion of the universe.

This lower radiated power permits the survival of much smaller black holes from the early universe to the present which have much higher temperatures and hence much higher energy photons. These have the potential of shedding light on the observed gamma-ray background. Little black holes may be able to account for the missing mass of the universe; and possibly even ball lightning, since they meet the criteria for ball lightning. Taken as a whole, these criteria argue against most other models.

## Acknowledgment

I wish to express appreciation to my good friends Yeong Kim and Ned Brit for valuable discussions, and Felipe Garcia for his help.

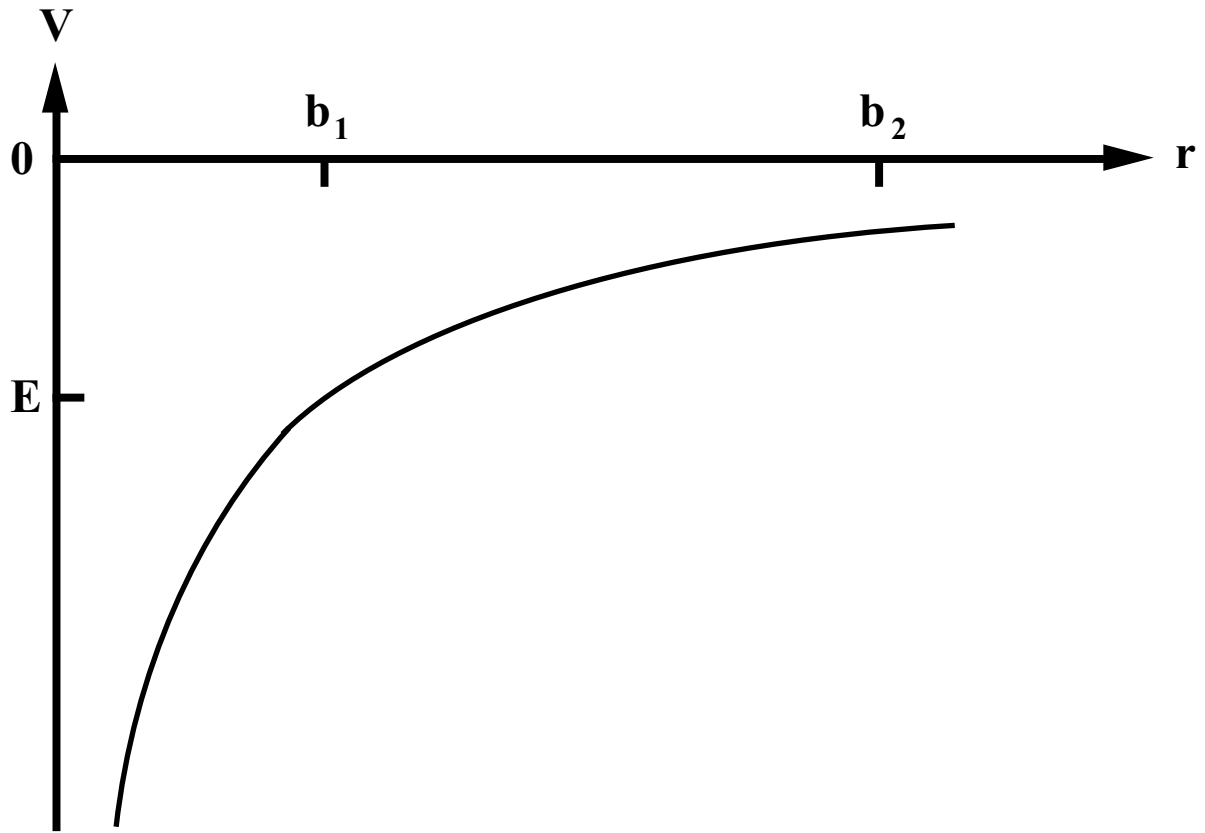

**Fig. 1.** Spherically symmetric gravitational potential energy of an isolated body.

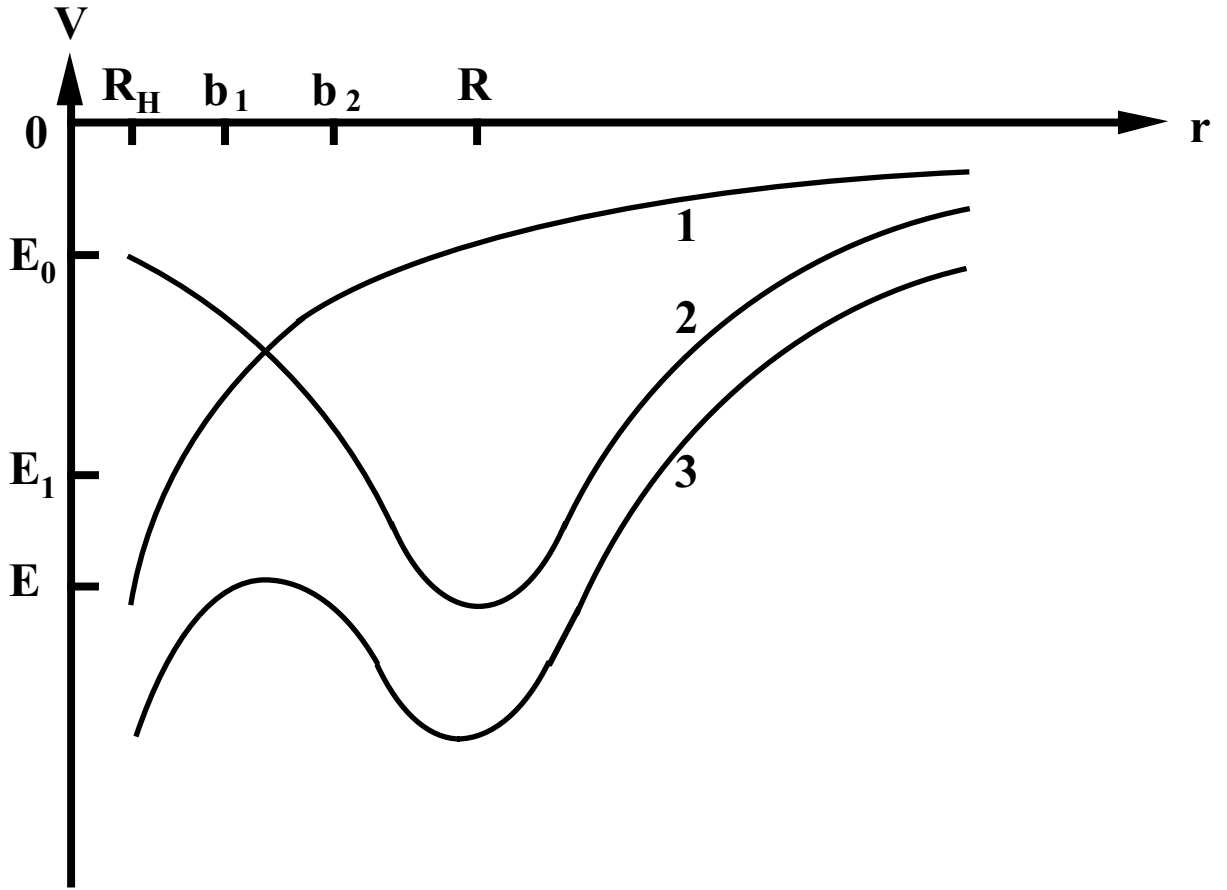

**Fig. 2.** Gravitational barrier resulting from mass $M_2$ at R opposite a black hole of mass M at the origin with Schwarzchild radius, $R_H$. The classical turning points are $b_1$ and $b_2$. Curve 1 is the effective potential of an isolated black hole. Curve 2 is the effective potential of mass $M_2$. Curve 3 is the effective potential of the two bodies.